%% file: paper.tex
\documentclass[11pt, notitlepage, tightenlines,aps, prd, reprint,superscriptaddress,nofootinbib]{revtex4-1}
\usepackage[utf8]{inputenc}

\usepackage{placeins}
\usepackage{mathtools}
\usepackage{amsfonts}
\usepackage{amssymb,amsmath}
\usepackage{color}
\usepackage[dvipsnames]{xcolor}
\definecolor{lcolor}{rgb}{0.5,0,0}
\definecolor{citcolor}{rgb}{0,0.3,0.0}
\usepackage[breaklinks,colorlinks,urlcolor=blue,citecolor=citcolor,linkcolor=lcolor]{hyperref}
\usepackage{graphicx}
\usepackage[caption=false]{subfig}
\graphicspath{ {image/} }
\usepackage{multirow}
\usepackage{booktabs}

\usepackage{braket}
\usepackage{bm}
\usepackage{txfonts}
\usepackage{mathrsfs} 
\usepackage{slashed}
\usepackage{accents}
\usepackage{hepunits}
\usepackage[normalem]{ulem}

\usepackage{soul} 
\usepackage{cancel} 

\usepackage{cleveref}
\crefname{section}{Sec.}{Secs.}
\crefname{figure}{Fig.}{Figs.}
\crefname{appendix}{App.}{Apps.}
\crefname{equation}{Eq.}{Eqs.}
\crefname{table}{Table}{Tables}
\Crefname{section}{Section}{Sections}
\Crefname{figure}{Figure}{Figures}
\Crefname{appendix}{Appendix}{Appendices}
\Crefname{equation}{Equation}{Equations}
\Crefname{table}{Table}{Tables}

\usepackage{tikz}
\usetikzlibrary{
    arrows.meta,
    decorations.pathmorphing,
    decorations.markings
}

\pgfdeclarelayer{background}
\pgfdeclarelayer{foreground}
\pgfsetlayers{background,main,foreground}

\def\p{{\boldsymbol p}}
\def\q{{\boldsymbol q}}
\def\k{{\boldsymbol k}}
\def\x{{\boldsymbol x}}
\def\y{{\boldsymbol y}}
\def\r{{\boldsymbol r}}

\def\l{{\boldsymbol l}}

\def\P{{\boldsymbol P}}
\def\TEEC {{\mathrm{TEEC}}}

\DeclareMathOperator{\Tr}{Tr}
\newcommand{\GeV}{{{\,}\textrm{GeV}}}

\begin{document}

\bibliographystyle{apsrev4-1}

\title{Heavy-flavor transverse energy-energy correlators in the color-glass condensate}

\author{Zhong-Bo Kang}
\email{zkang@physics.ucla.edu}
\affiliation{Department of Physics and Astronomy, University of California, Los Angeles, CA 90095, USA}
\affiliation{Mani L. Bhaumik Institute for Theoretical Physics,
University of California, Los Angeles, CA 90095, USA}
\affiliation{Center for Frontiers in Nuclear Science, Stony Brook University, Stony Brook, NY 11794, USA}

\author{Carlos Lamas}
\email{carloslamas.rodriguez@usc.es}
\affiliation{Instituto Galego de Fisica de Altas Enerxias (IGFAE), Universidade de Santiago de Compostela, E-15782 Galicia, Spain}

\author{Jani Penttala}
\email{jani.penttala@ijclab.in2p3.fr}
\affiliation{Universit\'e Paris-Saclay, CNRS/IN2P3, IJCLab, F-91405 Orsay, France}

\begin{abstract}

We calculate the transverse energy–energy correlator (TEEC) for forward heavy-quark pair production in proton–nucleus collisions at leading order within the Color Glass Condensate framework. 
The calculation retains the heavy-quark mass, exact final-state parton kinematics and full relative azimuthal-angle dependence, without imposing a back-to-back approximation. Within the adopted collinear fragmentation prescription, a complete sum over hadron species in the two heavy-quark fragmentation branches eliminates the fragmentation functions through the momentum sum rule. We present numerical results for charm production at RHIC energies and charm and bottom production at LHC energies, using HERA-constrained McLerran–Venugopalan initial conditions evolved with the running-coupling Balitsky–Kovchegov equation. Relative to proton–proton collisions, the nuclear target suppresses the integrated TEEC per nucleon and shifts the relative angular weight towards configurations farther from back-to-back. The magnitude and angular dependence of these modifications vary with the quark mass, collision kinematics and nuclear geometry. These results illustrate the sensitivity of heavy-flavor energy correlations to nuclear small-$x$ dynamics while reducing their leading collinear fragmentation dependence.

\end{abstract}

\maketitle

\section{Introduction} \label{sec:Introduction}

Understanding the behavior of quantum chromodynamics (QCD) at high parton densities is a central goal of high-energy nuclear physics. At sufficiently small longitudinal momentum fractions $x$, the growth
of the gluon density makes nonlinear effects and coherent multiple scattering essential, which can be described in the Color Glass Condensate (CGC) effective field theory approach~\cite{Gelis:2010nm,Weigert:2005us}.
The resulting dynamics is characterized by a saturation momentum $Q_s$, which generally increases towards smaller
$x$ and for larger nuclei. In the McLerran--Venugopalan (MV) model, the color field of a large nucleus is generated by an ensemble of color sources, allowing multiple scatterings to be treated to all orders in the target field~\cite{McLerran:1993ni,McLerran:1993ka,McLerran:1994vd}. The evolution of the target towards smaller $x$ is described by the JIMWLK renormalization-group equation~\cite{Jalilian-Marian:1996mkd,
Jalilian-Marian:1997qno,Iancu:2000hn,Ferreiro:2001qy}, while the Balitsky--Kovchegov (BK) equation provides a closed evolution equation for the dipole correlator within a large-$N_c$ mean-field approximation~\cite{Balitsky:1995ub,Kovchegov:1999yj}.

Forward particle production in proton--nucleus collisions offers access to this high-density regime by combining a relatively large-$x$ projectile parton with small-$x$ degrees of freedom in the target. Heavy-quark production is particularly useful in this context because it probes the small-$x$ gluon dynamics of the target, while the heavy-quark mass provides a perturbative scale for the production process. Quark--antiquark pair production in the CGC has been derived including multiple scattering and the interference between splitting before and after the target interaction~\cite{Blaizot:2004wv}. Its dependence on the quark mass, nuclear size and small-$x$ evolution has been investigated in both general studies of pair production and open-heavy-flavor phenomenology~\cite{Fujii:2006ab,Fujii:2013yja}.

Two-particle correlations provide information beyond inclusive yields by resolving the transverse-momentum imbalance generated by the target. In the correlation limit, in which this imbalance is small compared
with the individual hard momenta, the CGC description can be related to transverse-momentum-dependent gluon distributions~\cite{
Dominguez:2011wm}. Heavy-quark pairs are sensitive to both unpolarized and linearly polarized small-$x$ gluons~\cite{Marquet:2017xwy}, and their
production has also been studied within improved transverse-momentum factorization~\cite{Altinoluk:2021ygv}. A quantitative interpretation of the back-to-back region additionally requires soft-gluon radiation: Sudakov logarithms can be resummed together with small-$x$ evolution~\cite{Mueller:2013wwa}. Recent applications have studied forward heavy-meson pair correlations with Sudakov resummation~\cite{Gao:2026azd}, including a calculation over the full azimuthal range~\cite{Salazar:2026bmt}. These developments motivate complementary observables that retain sensitivity to the target while reducing the explicit dependence on fragmentation functions (FFs).

Energy correlators offer such a possibility by measuring angular correlations weighted by the energy carried by the detected particles. The energy--energy correlator (EEC) was originally introduced in
electron--positron annihilation~\cite{Basham:1978bw,Basham:1978zq}, and
its transverse counterpart, the transverse energy--energy correlator (TEEC), was subsequently proposed for hadron collisions~\cite{Ali:1984yp}. Energy weights and inclusive sums over final-state particles relate these observables to momentum flow and can simplify their dependence on collinear fragmentation through momentum sum rules. 
Their perturbative structure has been studied extensively, including
factorization and resummation for the back-to-back
TEEC~\cite{Gao:2019ojf,Gao:2023ivm}. In the collinear regime, the EEC
and its generalizations have been studied as probes of perturbative
jet evolution~\cite{Dixon:2019uzg,Chen:2020vvp,Jaarsma:2023ell} and
the transition from partons to hadrons~\cite{Komiske:2022enw,
Barata:2024wsu,Liu:2024lxy,Lee:2025okn,Kang:2025zto,Chang:2025kgq,Guo:2025zwb,Barata:2026uby,Fu:2024pic,Herrmann:2025fqy}. Measurements of energy
correlators inside jets have further provided experimental access
to these perturbative and nonperturbative
regimes~\cite{CMS:2024mlf,ALICE:2024dfl,ALICE:2026giw}.

Energy correlations have also emerged as probes of nuclear and small-$x$ dynamics. Nucleon energy correlators have been proposed as inclusive observables sensitive to gluon saturation~\cite{Liu:2023aqb}, and the TEEC has been studied in the CGC framework for electron--hadron scattering at the Electron--Ion Collider~\cite{
Kang:2023oqj}. In proton--nucleus collisions, photon--hadron and photon--jet TEECs have been investigated using the dipole description of the target, with fragmentation-function cancellation demonstrated
under the corresponding inclusive assumptions~\cite{Kang:2025vjk}.
Further photon--hadron studies have examined the effects of parton showers and hadronization~\cite{Ganguli:2025aqa}.
These observables probe different aspects of the angular distribution of energy and provide a basis for extending the study to heavy-flavor production.

In heavy-ion collisions, in-jet EECs probe medium-induced radiation
and color coherence~\cite{Andres:2022ovj,Andres:2023xwr}, and CMS has
observed their modification in lead--lead relative to proton--proton
collisions~\cite{CMS:2025ydi}. Studies of dijet TEECs also explore
transverse-momentum broadening, radiative corrections and medium
response in the back-to-back region~\cite{Barata:2026tis}.
These applications complement small-$x$ studies by probing
medium-induced changes in angular energy flow.

In this work, we investigate the hadron--hadron TEEC associated with forward heavy-quark pair production in proton--nucleus collisions. We consider correlations between hadrons in the two different heavy-quark fragmentation branches, sum over all hadron species in
each branch and integrate over their transverse momenta. The angular dependence is expressed through
$\tau=(1+\cos\phi)/2$, where $\phi$ is the relative azimuthal angle; the back-to-back and near-side limits correspond to $\tau\to0$ and $\tau\to1$, respectively. Within the adopted collinear fragmentation prescription and complete momentum sums, the transverse-momentum weights reduce the fragmentation convolutions to their first moments.
The momentum sum rules then eliminate the explicit collinear fragmentation functions. 

We evaluate the production process at leading order in the hard splitting, retaining the heavy-quark mass and the full relative azimuthal-angle dependence without imposing the correlation limit. The target interaction is treated eikonally, with a leading-color approximation for the required target correlators, while exact massive final-state parton kinematics is used at fixed pseudorapidities in the nucleon--nucleon center-of-mass frame. For the target, we use an MV-inspired dipole S-matrix with a HERA-constrained initial condition~\cite{Lappi:2013zma,Casuga:2023dcf}, evolved to smaller $x$ with the running-coupling BK equation~\cite{Balitsky:1995ub,Kovchegov:1999yj} using Balitsky's prescription~\cite{Balitsky:2006wa}. The projectile gluon distribution is taken from CT18~\cite{Hou:2019efy}, using LHAPDF for the parton distribution function (PDF) and running coupling~\cite{Buckley:2014ana}.

We present predictions for charm production at RHIC and for charm and bottom production at the LHC. We study the TEEC per target nucleon and the corresponding nuclear modification factor $R_{pA}$, examining how the nuclear target modifies the magnitude and angular dependence of the correlations. The comparison between proton and nuclear targets probes the sensitivity of these observables to the enhanced nuclear saturation scale and coherent multiple scattering. Our numerical study focuses on the interior of the angular range; a treatment including Sudakov resummation and a consistent description of the collinear region is left for future work.

The remainder of this paper is organized as follows. \cref{sec:framework} introduces the kinematics, derives the heavy-flavor TEEC and describes the target model and scale choices. \cref{sec:results} presents the numerical results and their dependence on the quark flavor and collision kinematics. \cref{sec:conclusions} summarizes our conclusions and discusses future extensions.
 
\section{Theoretical framework}\label{sec:framework}

In this section, we derive the analytic expression of the studied observable and explain the models and parametrizations used in the calculation. We start with an explanation of the considered kinematical setup. We then show how the TEEC is derived from the partonic cross section and its insensitivity to fragmentation functions. Finally, we explain the modeling of the interaction with the nuclear target and the PDF and scale choice.

\subsection{Kinematical setup} \label{sec:kinematics}

We consider proton--nucleus collisions, with the proton moving in the
positive longitudinal direction and the nucleus in the negative one.
Throughout this work, all calculated momenta, rapidities and
pseudorapidities are defined in the nucleon--nucleon center-of-mass
(CM) frame. We use light-cone coordinates $v^\pm = (v^0 \pm v^z)/\sqrt{2}$. Neglecting the beam masses, the incoming momenta are
\begin{equation}
    P^\mu = (P^+,0,\boldsymbol{0}),
    \quad
    P_A^\mu = (0,P_A^-,\boldsymbol{0}),
    \quad
    P^+ = P_A^- = \sqrt{ \frac{ s_{NN}}{2}},
\end{equation}
where $P^\mu$ is the proton momentum, and $P_A^\mu$ is the target momentum per nucleon.

The outgoing heavy quark and antiquark are on shell, with mass $m_Q$.
Their four-momenta can be expressed as
\begin{equation}
    p_i^\mu =
    \left(
        \frac{m_{T,i}}{\sqrt{2}}e^{y_i},
        \frac{m_{T,i}}{\sqrt{2}}e^{-y_i},
        \boldsymbol{p}_i
    \right),
    \qquad
    m_{T,i} = \sqrt{m_Q^2+p_{T,i}^2},
    \label{eq:kin_parton_momentum}
\end{equation}
where $y_i$ is the rapidity and $\boldsymbol{p}_i$ is the transverse momentum vector with magnitude $ p_{T,i}=|\boldsymbol{p}_i|$. It is convenient to express the observable in terms of
pseudorapidity,
\begin{equation}
    \eta_i = -\ln\tan\frac{\theta_i}{2}.
\end{equation}
At fixed transverse momentum, the longitudinal momentum and energy
are
\begin{equation}
    p_{z,i}=p_{T,i}\sinh\eta_i,
    \qquad
    E_i=\sqrt{m_Q^2+p_{T,i}^2\cosh^2\eta_i}.
    \label{eq:kin_energy_pseudorapidity}
\end{equation}
Consequently, the exact relation between rapidity and pseudorapidity
is
\begin{equation}
    y_i =
    \frac{1}{2}\ln\frac{E_i+p_{z,i}}{E_i-p_{z,i}}
    =
    \operatorname{arsinh}
    \left(
        \frac{p_{T,i}}{m_{T,i}}\sinh\eta_i
    \right).
    \label{eq:kin_exact_rapidity}
\end{equation}
The corresponding Jacobian is
\begin{equation}
    \mathcal{J}_i
    \equiv
        \frac{\partial y_i}{\partial\eta_i}
    =
    \frac{p_{T,i}\cosh\eta_i}{E_i}.
    \label{eq:kin_rapidity_jacobian}
\end{equation}
A cross section differential in the two pseudorapidities therefore
contains the factor $\mathcal{J}_1\mathcal{J}_2$. Since these
Jacobians depend on the transverse momenta, they must remain inside
the transverse-momentum integrations.

The longitudinal momentum fractions probed in the projectile and
target are determined by the outgoing pair. The projectile supplies the pair's plus momentum, while the target supplies its minus momentum. Thus,
\begin{align}
\begin{split}
    x_p &= \frac{p_1^++p_2^+}{P^+} = \frac{E_1+p_{z,1}+E_2+p_{z,2}}{\sqrt{s_{NN}}},
    \\
    x_t &= \frac{p_1^-+p_2^-}{P_A^-} = \frac{E_1-p_{z,1}+E_2-p_{z,2}}{\sqrt{s_{NN}}}.
    \label{eq:kin_exact_momentum_fractions}
\end{split}
\end{align}
with $E_i$ and $p_{z,i}$ expressed in terms of the transverse momentum and pseudorapidity as in \cref{eq:kin_energy_pseudorapidity}.
The incoming projectile gluon has plus momentum
$p^+=p_1^++p_2^+$, and the fraction carried by the quark is
\begin{equation}
    z =
    \frac{p_1^+}{p_1^++p_2^+}
    =
    \frac{E_1+p_{z,1}}
         {E_1+p_{z,1}+E_2+p_{z,2}}.
    \label{eq:kin_exact_splitting_fraction}
\end{equation}
The antiquark carries the complementary fraction $1-z$.
The transverse-momentum integrations are restricted to the physical
domain
\begin{equation}
    0<x_p<1,
    \qquad
    0<x_t< 1,
    \qquad
    0<z<1.
\end{equation}

For fragmentation, we adopt the spatial collinear prescription
\begin{equation} \label{eq:FragmentationPrescription}
    \vec{p}_{h_i}=\xi_i\vec{p}_i,
    \qquad 0<\xi_i<1,
\end{equation}
defined in the CM frame. Different choices can affect the cross section, especially at low $p_T$ \cite{Cacciari:2003uh}. 
With the prescription in Eq.~\eqref{eq:FragmentationPrescription}, the hadron and its parent parton are collinear, and
\begin{equation} \label{eq:FragmentationPrescriptionTransformation}
    \eta_{h_i}=\eta_i,
    \qquad
    \phi_{h_i}=\phi_i,
    \qquad
    p_{T,h_i}=\xi_i p_{T,i}.
\end{equation}
The equality of pseudorapidities holds exactly within this
prescription, including for massive particles; however, their rapidities
need not coincide.

\subsection{The transverse energy-energy correlator} \label{sec:TEEC}

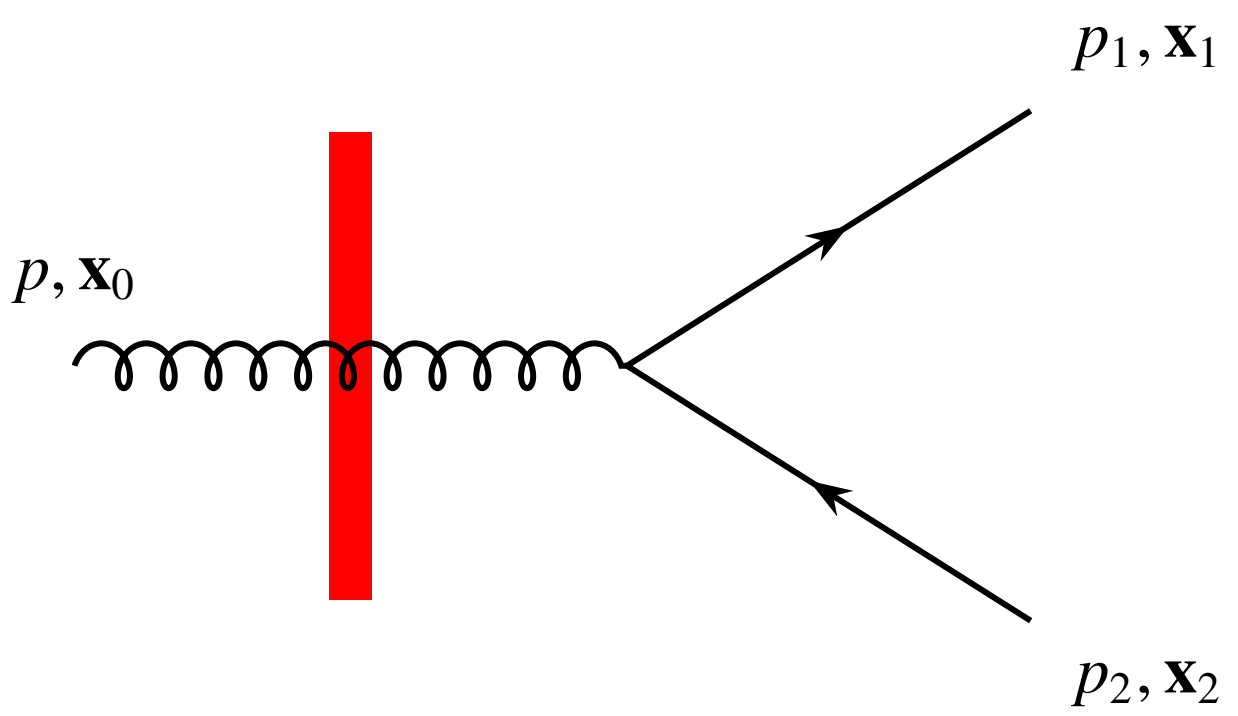
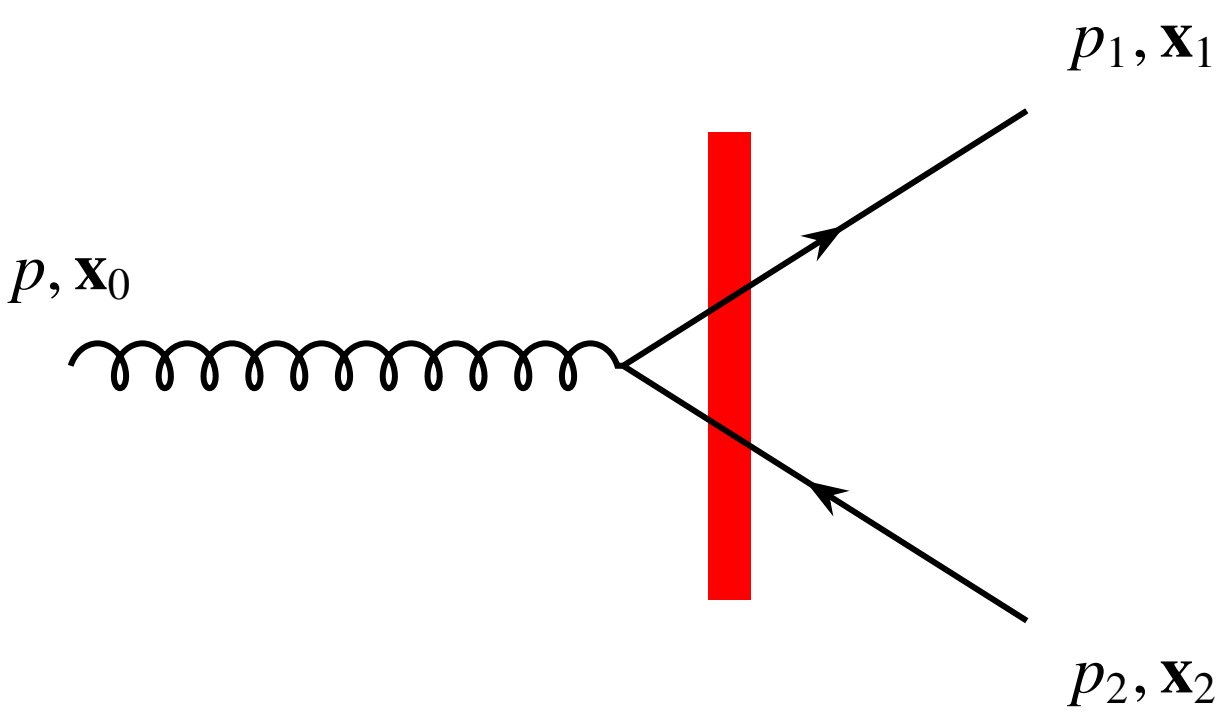
\begin{figure*}[t]
    \centering

    \subfloat[Splitting after the interaction%
        \label{fig:PartonicDiagram_a}]{%
        \resizebox{0.47\textwidth}{!}{%
            \input{g2QQbar_after}%
        }%
    }
    \hfill
    \subfloat[Splitting before the interaction%
        \label{fig:PartonicDiagram_b}]{%
        \resizebox{0.47\textwidth}{!}{%
            \input{g2QQbar_before}%
        }%
    }

    \caption{Parton-level diagrams for $Q\bar{Q}$ pair production
    in a $p+A$ collision at leading order in $\alpha_s$.
    The target shockwave that scatters the projectile is represented by the narrow red block. The splitting can happen after (left panel) or before (right panel) the shockwave interaction.}
    \label{fig:PartonicDiagrams}
\end{figure*}

At partonic level and lowest possible order in $\alpha_s$ there are two diagrams allowing for $Q\bar{Q}$ pair creation, shown in \cref{fig:PartonicDiagrams}. The squared amplitude associated with these diagrams is
\begin{align} \label{eq:MomentumTotalAmplitude}
    \langle |\mathcal{M}|^2 \rangle & = 2 g^2 (2\pi)^2 (2p^+)^2 z(1-z) S_\perp \int d^2 \l\, \mathcal{K}(\k, \l; m_Q, z) \notag \\
    & \times \langle \mathcal{F}_{x_t}(z\P + \k-\l) \mathcal{F}_{x_t}(-(1-z)\P + \k-\l) \rangle ,
\end{align}
where $\P = \p_1 + \p_2$ and $\k = (1-z) \p_1 - z\p_2$. Here $\p_1$ and $\p_2$ are the transverse momenta of quark and antiquark, respectively. Here $\mathcal{F}_{x_t}(\q)$ represents the momentum space dipole amplitude
\begin{align} \label{eq:MomentumSpaceDipole}
    \mathcal{F}_{x_t}(\q) = \int \frac{d^2\r}{(2\pi)^2} e^{-i\q \cdot \r} S_{x_t}(\x, \y) ,
\end{align}
with $\r = \x - \y$ and $S_{x_t}(\x, \y)$ being the dipole S-matrix for the interaction with the target
\begin{align}
    S_{x_t}(\x, \y) \equiv \frac{1}{N_c} \Tr [V(\x) V^\dagger(\y)]\, .
\end{align}
where $V(\x)$ is the Wilson line. We have also defined the splitting kernel
\begin{align} \label{eq:SplittingKernel}
    \mathcal{K}_{g\to Q\bar{Q}} (\k, \l; m_Q, z)  & = T_F m_Q^2 \left( \frac{1}{\k^2 + m_Q^2} - \frac{1}{\l^2 + m_Q^2} \right)^2 \notag \\
    & + P_{g\to q\bar{q}}(z) \left( \frac{\k}{\k^2 + m_Q^2} - \frac{\l}{\l^2 + m_Q^2} \right)^2 \, ,
\end{align}
with $P_{g \to q\bar{q}} (z) = T_F [z^2 + (1-z)^2]$ being the real contribution to the DGLAP kernel. This splitting kernel is equivalent to the one obtained in the framework of light-front perturbation theory in coordinate space in \cite{Marquet:2017xwy}. In the derivation of \cref{eq:MomentumTotalAmplitude} we have assumed that the target is translationally invariant and isotropic, so the dipole amplitude only depends on the separation between the particles. We also worked in the large-$N_c$ limit so the cross section can be expressed only in terms of dipole amplitudes and quadrupoles would only appear at $\mathcal{O}(1/N_c^2)$ which is not considered here. Note that the process is not possible in vacuum, so the non-interacting term of the S-matrix will not contribute to the cross section.

From the amplitude it is easy to obtain the quark-antiquark production cross section
\begin{align} \label{eq:GeneralCrossSection}
    d\sigma = 2\pi \delta(p^+-p_1^+ & -p_2^+) \notag \\
    & \times \frac{1}{2p^+} \frac{dp_1^+}{2p_1^+} \frac{d^2 \p_1}{(2\pi)^3} \frac{dp_2^+}{2p_2^+} \frac{d^2 \p_2}{(2\pi)^3} \langle|\mathcal{M}|^2\rangle\, .
\end{align}
We are interested in the cross section in terms of the particle pseudorapidities and transverse momenta. The parton level cross section then follows from \cref{eq:MomentumTotalAmplitude} as
\begin{align} \label{eq:PartonicCrossSection}
    \frac{d\sigma^{gA\to Q\bar{Q}}}{d\eta_1 d\eta_2 d^2\p_1 d^2\p_2} & = \frac{2\alpha_s(\mu_R^2)}{(2\pi)^2} S_\perp\, \delta(p^+ - p^+_1 - p^+_2) p^+ \notag \\
    & \times \mathcal{J}_1 \mathcal{J}_2\,  z (1-z) \mathcal{I}_{x_t}(\p_1, \p_2; m_Q, z) ,
\end{align} 
where $\mathcal{J}_i$ is the transformation Jacobian from rapidity to pseudorapidity, as defined in \cref{eq:kin_rapidity_jacobian}. We have defined the interaction kernel
\begin{align} \label{eq:InteractionKernel}
    \mathcal{I}_{x_t}(\p_1, \p_2 & ; m_Q, z)
     \equiv \int d^2 \l\, \mathcal{K}_{g\to Q\bar{Q}} (\k, \l; m_Q, z) \notag \\
    & \times \langle \mathcal{F}_{x_t}(z\P + \k-\l) \mathcal{F}_{x_t}(-(1-z) \P + \k-\l) \rangle
    \, .
\end{align}
To obtain the proton-nucleus cross section, the partonic cross section must be convoluted with the gluon PDF inside the proton and the FF of the heavy quarks into the final hadrons
\begin{widetext}
\begin{align} \label{eq:HadronConvolution}
    \frac{d\sigma^{pA \to h_1h_2}}{d\eta_{h_1} d\eta_{h_2} d^2\p_{h_1} d^2\p_{h_2}} = \int \frac{dp^+}{p^+} x_p f_{g/p}(x_p, \mu)\int_{0}^{1} d\xi_1 & D_{h_1/Q}(\xi_1, \mu) \notag \\
    & \times \int_{0}^{1} d\xi_2 D_{h_2/\bar{Q}}(\xi_2, \mu) \frac{d\sigma^{gA\to Q\bar{Q}}}{d\eta_1 d\eta_2 d^2\p_1 d^2\p_2} \left| \frac{\partial \p_1}{\partial \p_{h_1}} \right| \left| \frac{\partial \p_2}{\partial \p_{h_2}} \right| \frac{\partial \eta_1}{\partial \eta_{h_1}} \frac{\partial \eta_2}{\partial \eta_{h_2}}\, ,
\end{align}
\end{widetext}
where $\mu$ is the factorization scale and $\xi_i$ is the fraction of the heavy quark momenta carried by the outgoing hadron. Within the fragmentation prescription introduced in \cref{eq:FragmentationPrescription} and \cref{eq:FragmentationPrescriptionTransformation} it is immediate to see that $|\partial \p_i / \partial \p_{h_i}| = 1/\xi_i^2$ and $\partial \eta_i / \partial \eta_{h_i} = 1$. 
The di-hadron production cross section then reads
\begin{align} \label{eq:DihadronCrossSect}
    \frac{d\sigma^{pA \to h_1h_2}}{d\eta_{h_1} d\eta_{h_2} d^2\p_{h_1} d^2\p_{h_2}} & = \frac{2\alpha_s(\mu_R^2)}{(2\pi)^2} S_\perp \int_{0}^{1} \frac{d\xi_1}{\xi_1^2}  D_{h_1/Q}(\xi_1, \mu) \notag \\
    & \hspace{-40 pt} \times \int_{0}^{1} \frac{d\xi_2}{\xi_2^2} D_{h_2/\bar{Q}}(\xi_2, \mu) x_p f_{g/p}(x_p, \mu) \notag \\ 
    & \hspace{-40 pt} \times \mathcal{J}_1 \mathcal{J}_2 z (1-z) \mathcal{I}_{x_t} (\p_1, \p_2; m_Q, z)\, . 
\end{align}

We are interested in studying correlations between the final state hadrons as a function of the azimuthal separation between them. Weighting the contribution of the different hadrons by their transverse momentum helps reduce sensitivity to fragmentation and reduces theoretical uncertainties. We here focus on the two-point function, the transverse energy-energy correlator, that measures the correlations between the transverse momentum flowing into two different directions of the transverse plane. We sum over all the fragmentation products in the two heavy-quark branches to obtain
\begin{align}
    \frac{d \Sigma^{h+h}}{d\tau d\eta_{h_1} d\eta_{h_2}} & = \sum_{h_1 h_2} \int d^2\p_{h_1} d^2\p_{h_2} \delta \left( \tau - \frac{1+\cos\phi}{2} \right) \notag \\
    & \times \frac{d\sigma^{pA \to h_1h_2}}{d\eta_{h_1} d\eta_{h_2} d^2\p_{h_1} d^2\p_{h_2}} \frac{p_{T, h_1} p_{T, h_2}}{(\sum_{h^\prime} p_{T, h^\prime})^2}\, .
\end{align}
where the sum runs over all the products of the quark-antiquark pair hadronization.
We only consider correlations between two particles coming from different branches. The case where the two particles come from the fragmentation of the same parton, which would sit exactly at $\tau=1$, is excluded. We only count each cross-branch pair once.
We want to rewrite this formula in terms of the parton momenta. We know that $d^2 \p_{h_i} = \xi_i^2 d^2 \p_i$ and $d\eta_{h_i} = d\eta_i$. The energy weight yields 
\begin{align}
    \frac{p_{T, h_1} p_{T, h_2}}{(\sum_{h^\prime} p_{T, h^\prime})^2} = \xi_1 \xi_2 \frac{p_{T,1} p_{T,2}}{(p_{T,1} + p_{T,2})^2} \, .
\end{align}
Putting everything together and using the momentum sum rule
\begin{align} \label{eq:FragmentationSumRule}
    \sum_{h} \int d\xi \xi D_{h/Q}(\xi, \mu) = 1\, ,
\end{align}
we get rid of the fragmentation functions within our collinear fragmentation prescription. The transverse energy-energy correlator, normalized by the number of nucleons, then follows as
\begin{widetext}
\begin{align} \label{eq:TEEC}
    \TEEC^{h+h}  \equiv \frac{1}{A} \frac{d \Sigma^{h+h}}{d\tau d\eta_1 d\eta_2} = \int d^2 \p_1 d^2 \p_2\frac{2 \alpha_s(\mu_R^2)}{(2\pi)^2}  \frac{S_\perp}{A} & z(1-z) x_p f_{g/p}(x_p,\mu) \notag \\
    & \times \mathcal{J}_1 \mathcal{J}_2 \frac{p_{T, 1} p_{T, 2}}{(p_{T,1} + p_{T,2})^2} \mathcal{I}_{x_t}(\p_1, \p_2; m_Q, z) \delta \left( \tau - \frac{1+\cos\phi}{2} \right) \, .
\end{align}
\end{widetext}
Within the adopted collinear fragmentation prescription and complete
sums over hadron species and momenta, the explicit fragmentation
functions cancel from the TEEC through the momentum sum rule.
This reduces fragmentation-related model dependence, providing
cleaner access to the production dynamics and the target's small-$x$
gluon structure, and facilitating comparisons with experimental data.
With our definition of the angular
variable, $\tau\to0$ corresponds to the back-to-back region and
$\tau\to1$ to the near-side region.

To address the effect of gluon saturation on the observable it is useful to also introduce the nuclear modification factor
\begin{align} \label{eq:Rpa}
    R_{pA}^\Sigma = \frac{\TEEC_{pA}^{h+h}}{\TEEC_{pp}^{h+h}}.
\end{align}
Normalizing the nuclear target TEEC by the proton-proton result, where saturation effects are expected to be weaker, one can better identify how the TEEC is modified by saturation effects.

\subsection{The dipole amplitude} \label{sec:MV}

We need to specify a model to account for the interaction with the target. If the partons from the target carry a small fraction of the total nucleon momentum we can use the Color Glass Condensate (CGC) effective field theory to study the dynamics of the partons inside the high-energy nucleus.
In particular, we will use the McLerran-Venugopalan (MV) model, where the relevant physical scale is the saturation scale $Q_s$, which sets the transverse momentum scale below which nonlinear effects slow the growth of the gluon density, leading to gluon saturation. For this model, and in the large-$N_c$ limit, the expectation value of the product of dipole amplitudes factorizes into the product of expectation values \cite{Lappi:2015vta}
\begin{align}
    \langle S_{x_t}(\r) S_{x_t}(\tilde{\r}) \rangle \simeq \langle S_{x_t}(\r) \rangle \langle S_{x_t}(\tilde{\r}) \rangle ,
\end{align}
and the same for the momentum space dipole amplitudes. To obtain results for the TEEC we need to know $\langle S_{x_t}(r)\rangle$ for the probed $x_t$ values and Fourier transform them to momentum space to obtain $\langle \mathcal{F}_{x_t}(q)\rangle$.

We start from an MV-based functional form, where the dipole amplitude reads \cite{Lappi:2013zma}
\begin{align} \label{eq:MVeModel}
    \langle S_{x_t}(r) \rangle = \exp\left\{-\frac{(r^2 Q_{s,0}^2)^\gamma}{4} \log\left( \frac{1}{r\Lambda} + e \cdot e_c \right)\right\} .
\end{align}
This dipole amplitude is taken as an initial condition at a given value of the momentum fraction $x_t$ and then evolved to the desired value of $x_t$ through the Balitsky-Kovchegov (BK) evolution equation 
\begin{align} \label{eq:BK}
    \frac{\partial \langle S_{x_t}(r)\rangle}{\partial \log(1/x_t)} & = \int d^2 \r^\prime K^{run} (\r, \r^\prime) \notag \\
    & \times [\langle S_{x_t}(r^\prime) \rangle \langle S_{x_t}(|\r-\r^\prime|)\rangle - \langle S_{x_t}(r) \rangle] ,
\end{align}
where the BK evolution kernel in the Balitsky prescription~\cite{Balitsky:2006wa} reads
\begin{align}
    K^{run}(\r, \r^\prime) & = \frac{N_c \alpha_s(\r^2)}{2\pi^2} \left[ \frac{\r^2}{\r^{\prime\, 2} (\r - \r^\prime)^2} \right. \notag \\
    & + \frac{1}{\r^{\prime 2}} \left( \frac{\alpha_s(\r^{\prime2})}{\alpha_s((\r - \r^\prime)^2)} - 1 \right) \notag \\
    & \left. + \frac{1}{(\r - \r^\prime)^2} \left( \frac{\alpha_s((\r - \r^\prime)^2)}{\alpha_s(\r^{\prime 2})} - 1 \right) \right] .
\end{align}
Here, the strong coupling constant $\alpha_s$ is obtained through the coordinate-space running coupling
\begin{align}
    \alpha_s (r^2) = \frac{12 \pi}{(33- 2N_f) \log(4C^2/(r^2 \Lambda^2))}.
\end{align}
where $N_f = 3$, $\Lambda$ is the QCD scale parameter and $C^2$ is a parameter regulating the dependence of the coupling on $r$. In the infrared region the coupling is frozen to $\alpha_s = 0.7$. 

The initial values of the parameters in \cref{eq:MVeModel} are fitted to experimental data for proton targets. We here take the central values from the fit in \cite{Casuga:2023dcf}, corresponding to $\gamma=1$ for the initial condition. Namely, for $x_0 = 0.01$ the median values on the fit are $Q_{s0}^2 = 0.061 \GeV^2$, $\Lambda = 0.241 \GeV$, $e_c = 35.3$ and $C^2 = 4.97$ and $S_{\perp} = \sigma_0/2 \simeq 36.21 \GeV^{-2}$. For nuclear collisions this initial condition is scaled geometrically, considering that the saturation scale and target transverse area change with the nuclear mass number as~\cite{Kang:2023oqj}
\begin{align} \label{eq:NuclearScaling}
    Q_{s0, A}^2 = cA^{1/3}Q_{s0}^2\, , \qquad S_{\perp, A} = A^{2/3} S_\perp/c \, .
\end{align}
The dipole amplitude is then evolved with \cref{eq:BK} to obtain the interaction S-matrix at the $x_t$ value of the physical process.

\subsection{The parton distribution function and running coupling} \label{sec:PDF}

For phenomenological applications, we will use the CT18NLO proton PDF set \cite{Hou:2019efy} obtained through the LHAPDF library \cite{Buckley:2014ana}. The running coupling for the gluon splitting will also be obtained from the LHAPDF library, fixing the renormalization scale to be equal to the factorization scale and taking both to be the typical momentum scale of the process 
\begin{align} \label{eq:Scales}
    \mu^2 = \mu_R^2 = m_Q^2 + (\p_1^2 + \p_2^2)/2. 
\end{align}
With our scale prescription the PDF and the coupling depend on the final particle momenta so they must remain inside the transverse momentum integration.

\section{Results and discussion}\label{sec:results}

We present results\footnote{The code used for the numerical integration and plots is publicly available at \url{https://github.com/CLamas-r/HeavyQuark_TEEC.git}} for the TEEC and the nuclear modification factor for RHIC and LHC kinematics for charm ($m_c = 1.5\ \GeV$) and bottom ($m_b = 4.75\ \GeV$) quarks. For RHIC, the benchmark is a proton-gold collision with nucleon-nucleon center-of-mass energy of $\sqrt{s_{NN}} \simeq 200\ \GeV$. We evaluate the results at $\eta_1 = \eta_2 = 3.5$ inside the STAR forward tracking and calorimetry acceptance \cite{Brandenburg:2024syv}. For the LHC the benchmark is a proton-lead collision at $\sqrt{s_{NN}} \simeq 5.02\ \mathrm{TeV}$. We consider the TEEC at $\eta_1 = \eta_2 = 4$, within the LHCb nominal acceptance \cite{LHCb:2014set}. The factorization and renormalization scales are fixed as in \cref{eq:Scales} and the CT18NLO PDF and running coupling are obtained through the LHAPDF library. The interaction with the target is treated within the MV model as explained in \cref{sec:MV}.

The dipole amplitude model introduced in \cref{sec:MV} is only valid for the small-$x$ degrees of freedom of the target, and we must therefore make sure that the dominant contribution to the process comes from the small-$x$ partons. 
We consider our target description to be valid up to $x_t = 0.01$. To estimate the probed values of $x_t$ one can compute the average $\langle x_t \rangle$ weighted by their contribution to the TEEC.
At fixed pseudorapidities, we define the TEEC-weighted distribution
of target momentum fractions as
\begin{align}
    \frac{d\Sigma}{dx\,d\eta_1d\eta_2}
    & = \int d^2\p_1\,d^2\p_2\,
    \frac{d\sigma^{pA\to Q\bar Q}}
         {d\eta_1d\eta_2d^2\p_1d^2\p_2}
    \notag\\
    & \times
    \frac{p_{T,1}p_{T,2}}{(p_{T,1}+p_{T,2})^2}
    \delta\!\left(x-x_t(p_{T,1},p_{T,2};\eta_1,\eta_2)\right).
\end{align}
The corresponding average is
\begin{align} \label{eq:xtAverage}
    \langle x_t\rangle_\Sigma
    =
    \frac{\displaystyle\int_0^1 dx\,x\,
          \frac{d\Sigma}{dx\,d\eta_1d\eta_2}}
         {\displaystyle\int_0^1 dx\,
          \frac{d\Sigma}{dx\,d\eta_1d\eta_2}}.
\end{align}
The results are collected in \cref{tab:mean_xt} for the different targets both for charm and bottom quarks and RHIC and LHC kinematics. For the LHC kinematics we are deep into the small-$x$ region, where the target description is valid. For RHIC kinematics, where the center-of-mass energy is lower, we are closer to the upper edge of the small-$x$ range: For charm quark production we are still within the small-$x$ regime, but for bottom quarks $\langle x_t \rangle \sim 0.01$ and the observable is very sensitive to partons outside of the valid kinematic range. For this reason, we will only study charm quark production at RHIC and consider both charm and bottom production at LHC kinematics.
We have checked numerically that
among the benchmarks considered, only bottom production at RHIC
shows appreciable sensitivity to the large-$x_t$ region.

\begin{table*}[t]
    \centering
    \renewcommand{\arraystretch}{1.3}
    \setlength{\tabcolsep}{12pt}
    \setlength{\arrayrulewidth}{0.5pt}

    \begin{tabular}{l|cc|cc}
        & \multicolumn{2}{c|}{LHC}
        & \multicolumn{2}{c}{RHIC} \\
        Target & Charm & Bottom & Charm & Bottom \\
        \hline
        Proton
        & $2.9\times10^{-5}$ & $9.5\times10^{-5}$
        & $1.4\times10^{-3}$ & $9.5\times10^{-3}$ \\
        Nucleus, $c=0.5$
        & $2.9\times10^{-5}$ & $9.3\times10^{-5}$
        & $1.3\times10^{-3}$ & $8.9\times10^{-3}$ \\
        Nucleus, $c=1$
        & $3.0\times10^{-5}$ & $9.2\times10^{-5}$
        & $1.3\times10^{-3}$ & $8.5\times10^{-3}$ \\
    \end{tabular}
    \caption{TEEC-weighted average target momentum fraction
    $\langle x_t\rangle_{\Sigma}$ as defined in \cref{eq:xtAverage}. Results are shown for charm and bottom production at LHC and RHIC kinematics for the different target configurations.}
    \label{tab:mean_xt}
\end{table*}

In the collinear and back-to-back regions the contribution of large logarithms, i.e. Sudakov radiation in the back-to-back limit \cite{Mueller:2013wwa, Gao:2026azd} or collinear logarithms in the collinear limit~\cite{Dixon:2019uzg}, can become relevant.
We restrict the displayed results to $0.1\leq\tau\leq0.9$, avoiding
the angular endpoints where additional resummation or matching
treatments may be required. The quantitative impact of these
corrections on the massive observable is left for future work.

\begin{figure}[tb]
    \centering
    \includegraphics[width=\linewidth]{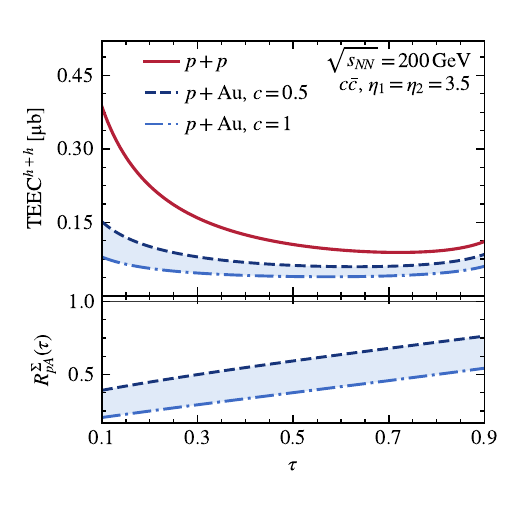}
    \caption{Transverse energy-energy correlator (TEEC), as defined in \cref{eq:TEEC}, normalized by the number of nucleons in the target (top panel) and nuclear modification factor, as defined in \cref{eq:Rpa}, (lower panel) for charm quarks at RHIC kinematics. The result for $pp$ collisions is represented by a red solid line. The result for the limiting values of the nuclear geometry uncertainty factor $c$ is represented by the dashed blue lines and the colored area between them represents the geometry dependence of the $pA$ results.}
    \label{fig:TEEC_RHIC}
\end{figure}

\begin{figure*}[tp]
    \centering
    \subfloat[Charm quarks\label{fig:TEEC_Charm_LHC}]{%
        \includegraphics[width=0.48\textwidth]{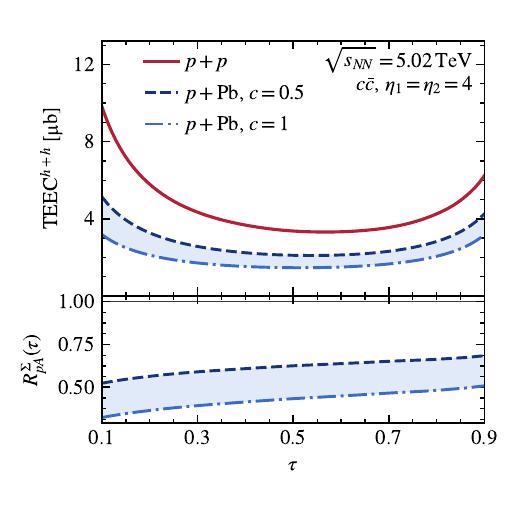}%
    }
    \hfill
    \subfloat[Bottom quarks\label{fig:TEEC_bottom_LHC}]{%
        \includegraphics[width=0.48\textwidth]{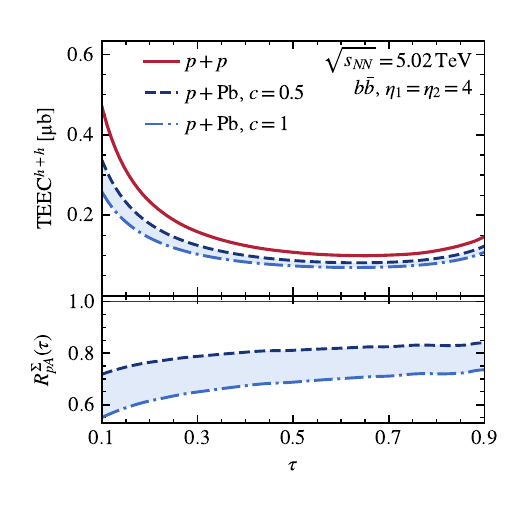}%
    }

    \caption{Transverse energy-energy correlator (TEEC), as defined in \cref{eq:TEEC}, normalized by the number of nucleons in the target (top panels) and nuclear modification factor, as defined in \cref{eq:Rpa}, (lower panels) for charm (left plot) and bottom (right plot) quarks at LHC kinematics. The results for $pp$ collisions are represented by a red solid line. The results for the limiting values of the nuclear geometry uncertainty factor $c$ are represented by the dashed blue lines and the colored area between them represents the geometry dependence of the $pA$ results.
    }
    \label{fig:TEEC_LHC}
\end{figure*}

The results for the TEEC and the nuclear modification factor are shown in \cref{fig:TEEC_RHIC} for the RHIC kinematics and in \cref{fig:TEEC_LHC} for the LHC kinematics. In all the cases we find strong nuclear effects with $R_{pA}$ generally far from unity. 
The increase of $R_{pA}^{\Sigma}$ with $\tau$ indicates that the
nuclear target enhances the relative weight of configurations farther
from back-to-back compared with the proton target. Within our
leading-order calculation, the incoming projectile gluon has zero
transverse momentum, and the pair's total transverse momentum is
supplied by the target. The angular redistribution is therefore
consistent with the stronger transverse-momentum broadening associated
with the enhanced nuclear saturation scale.

In addition to this redistribution of the angular correlation, the nuclear field also produces a suppression in the total weighted yield per nucleon.
In our geometry prescription, $S_{\perp,A}/A=S_{\perp,p}/(cA^{1/3})$. Consequently, writing the integrated TEEC at fixed pseudorapidities as \(\Sigma_T=S_{\perp,T}\mathcal W_T /A_T\), where \(\mathcal W_T\) is the corresponding weighted cross section per unit transverse area, gives
\begin{align}
\frac{\Sigma_A}{\Sigma_p}
=\frac{\mathcal W_A}{cA^{1/3}\mathcal W_p}.
\end{align}
The observed suppression therefore reflects an increase of the weighted production rate per unit area that is smaller than the nuclear thickness enhancement $cA^{1/3}$. 

In the dilute single-scattering approximation, the momentum-space dipole distribution can be expanded as
\begin{align}
\mathcal F_T(\q)
=\delta^{(2)}(\q)+f_T^{(1)}(\q)+\cdots,
\end{align}
where \(f_T^{(1)}\) is linear in the target density and the omitted terms contain higher powers of that density. For the initial \(\gamma=1\) MV model, its large-\(q\) behavior is
\begin{align}
f_T^{(1)}(\q)
\simeq \frac{Q_{s0,T}^{2}}{2\pi q^4}.
\end{align}
Keeping the terms linear in the target density in the product of dipole distributions, and using \(Q_{s0,A}^{2}=cA^{1/3}Q_{s0,p}^{2}\), gives
\begin{align}
\mathcal W_A^{\mathrm{dil}}
=cA^{1/3}\mathcal W_p^{\mathrm{dil}},
\qquad
\frac{\Sigma_A^{\mathrm{dil}}}{\Sigma_p^{\mathrm{dil}}}=1.
\end{align}
This proportionality is preserved if both targets subsequently undergo the same linear evolution. Coherent multiple scattering and nonlinear small-\(x\) evolution modify this linear response. The resulting suppression
depends on how these target effects are weighted by the heavy-quark
production kernel, the projectile gluon distribution and the TEEC weight
over the available phase space.

An important contribution arises from the change in the projectile
momentum fractions sampled by the observable. For our benchmarks, the
nuclear target shifts the TEEC-weighted transverse-momentum distribution
towards larger momenta compared with the proton target. At fixed
pseudorapidities, these configurations require larger values of $x_p$,
as follows from \cref{eq:kin_exact_momentum_fractions}.
This dependence is illustrated by the forward
estimate
\begin{align}
    x_p \simeq \frac{p_{T,1}e^{\eta_1}+p_{T,2}e^{\eta_2}}{\sqrt{s_{NN}}}.
\end{align}
The decrease of the projectile gluon distribution
$x_p f_{g/p}(x_p,\mu)$ over the relevant momentum-fraction range
penalizes these configurations and amplifies the suppression of the
nuclear yield. Although the projectile PDF is common to $pp$ and $pA$,
it weights different transverse-momentum distributions and therefore
does not cancel after integration.

This PDF reweighting is particularly relevant at RHIC, where the
observable probes larger projectile momentum fractions and the gluon
distribution falls more steeply. The lower collision energy also makes
the restriction $x_p<1$ more restrictive for contributions at larger
transverse momenta. Consequently, the stronger yield suppression found
at RHIC does not by itself imply a larger saturation scale than at the
LHC. Its magnitude reflects the combined effects of the target dynamics,
the projectile PDF and the available phase space.

The TEEC exhibits a similar qualitative pattern across quark flavors and collider kinematics, although the relative angular weights differ. At LHC kinematics, shown in \cref{fig:TEEC_LHC}, the near-side contribution relative to the back-to-back side of the displayed interval is less pronounced for bottom than for charm quarks. This trend is consistent with the mass regulation of nearly collinear \(g\to Q\bar Q\) splitting, reflected in the denominators \(k^2+m_Q^2\) of the production kernel. For charm production, the near-side rise is also weaker at RHIC than at the LHC, as shown in \cref{fig:TEEC_RHIC,fig:TEEC_LHC}. Changing the collision energy modifies the projectile and target momentum fractions and the available transverse-momentum phase space. These changes alter the relative contributions of different momentum configurations through the convolution of the massive production kernel, the projectile PDF and the target dipole distributions.

\section{Conclusion and outlook}\label{sec:conclusions}

We have studied transverse energy–energy correlations associated with forward heavy-quark pair production in proton–nucleus collisions within the Color Glass Condensate framework. We derived the leading-order TEEC, retaining the heavy-quark mass and the full relative azimuthal-angle dependence, without imposing a back-to-back approximation. The calculation includes the interference between gluon splitting before and after the shockwave interaction and uses exact final-state parton kinematics at fixed pseudorapidities. Within the adopted collinear fragmentation prescription, complete sums over the hadron species and momenta in the two fragmentation branches eliminate the dependence on collinear fragmentation functions through momentum sum rules.

Using an MV-inspired dipole S-matrix with a HERA-constrained initial condition, evolved to smaller $x$ with the running-coupling BK equation, we have presented predictions for charm production at RHIC and for charm and bottom production at the LHC. For the configurations considered, the nuclear target suppresses the TEEC yield per nucleon and shifts the relative angular weight towards configurations farther from back-to-back. This angular redistribution is consistent with transverse-momentum broadening generated by coherent multiple scattering in the nuclear color field, with the enhanced nuclear saturation momentum setting the characteristic scale of the broadening.

The yield suppression reflects the convolution of these target dynamics with the projectile gluon distribution and the available phase space. In particular, the nuclear target shifts the TEEC-weighted transverse momentum distribution towards larger values and hence larger projectile momentum fractions $x_p$, where the falling gluon PDF further suppresses the contributions. This amplification is particularly relevant at RHIC and contributes to the collision-energy dependence of the suppression. The dependence on the quark mass and collision kinematics therefore reflects the interplay of the production kernel, the projectile PDF and the target momentum distribution. These results motivate heavy-flavor TEECs as complementary probes of nuclear small-$x$ dynamics with reduced dependence on collinear fragmentation.

An important extension is the calculation of the hadron–hadron TEEC for massless partons. The massless limit exposes a collinear divergence in the contribution where the incoming gluon splits before interacting with the shockwave~\cite{Lappi:2012nh}. Subsequent scattering can deflect the daughter partons into different directions, so excluding the collinear region of the measured angular distribution does not remove this divergence. A consistent treatment therefore requires a factorization and subtraction prescription for the unresolved initial splitting, including its relation to multiparton scattering contributions. Developing this treatment will allow us to extend the calculation to light-parton channels and compare their nuclear modifications with those of heavy flavor.

We also plan to incorporate Sudakov resummation for the small-imbalance back-to-back region, consistently with the small-\(x\) evolution of the target. A complementary treatment of the near-side region will address final-state collinear dynamics, retaining the mass dependence and identifying the logarithms that require resummation in the relevant scale hierarchies. Matching these descriptions to the full-angle calculation while avoiding double counting will provide predictions over a broader angular range and permit a systematic comparison of heavy- and light-flavor energy correlations.

\section*{Acknowledgments}

We are grateful to Meijian Li for her help with the numerical implementation of the Fourier transform of the BK-evolved dipole amplitude. Z.K. is supported by the National Science Foundation under grant No.~PHY-2515057. C.L. is supported by Maria de Maeztu excellence unit grant CEX2023-001318-M, project PID2023-152762NB-I00 and predoctoral fellowship PRE2022-102748 funded by MICIU/AEI/10.13039/501100011033; by ERDF/EU; and by Xunta de Galicia project ED431F 2023/10.L and CIGUS Network of Research Centres. J.P. is supported by the UPSaclay STAR $\varphi$ programme that has received funding from the European Union's COFUND action, a part of the Marie Sklodowska-Curie Actions Program within the European Commission MSCA framework, UPSaclay STAR $\varphi$ Grant Agreement ID: 101216532, co-funded by the European Union. Views and opinions expressed are however those of the author(s) only and do not necessarily reflect those of the European Union or European Research Executive Agency. Neither the European Union nor the REA can be held responsible for them.

\appendix

\bibliography{bibliography}

\end{document}

%% file: g2QQbar_after.tex
\begin{tikzpicture}[
    line width=0.8pt,
    gluon line/.style={
        decorate,
        decoration={
            coil, aspect=0.7,
            segment length=6pt, amplitude=3pt
        }
    },
    fermion line/.style={
        decoration={
            markings,
            mark=at position 0.55 with {
                \arrow{Stealth[length=2mm,width=1.4mm]}
            }
        },
        postaction={decorate}
    }
]
    \coordinate (incoming)   at (0,0);
    \coordinate (splitting)  at (2.6,0);
    \coordinate (quark)      at (4.5,1.2);
    \coordinate (antiquark)  at (4.5,-1.2);

    \fill[red] (1.2,-1.1) rectangle (1.4,1.1);

    \draw[gluon line]   (incoming) -- (splitting);
    \draw[fermion line] (splitting) -- (quark);
    \draw[fermion line] (antiquark) -- (splitting);

    \node[above=5pt] at (incoming)
        {$p,\mathbf{x}_0$};
    \node[above right=2pt] at (quark)
        {$p_1,\mathbf{x}_1$};
    \node[below right=2pt] at (antiquark)
        {$p_2,\mathbf{x}_2$};
\end{tikzpicture}

%% file: g2QQbar_before.tex
\begin{tikzpicture}[
    line width=0.8pt,
    gluon line/.style={
        decorate,
        decoration={coil,aspect=0.7,
                    segment length=6pt,amplitude=3pt}
    },
    fermion line/.style={
        decoration={markings,
            mark=at position 0.55 with {
                \arrow{Stealth[length=2mm,width=1.4mm]}
            }
        },
        postaction={decorate}
    }
]
    \coordinate (incoming) at (0,0);
    \coordinate (splitting) at (2.6,0);
    \coordinate (quark) at (4.5,1.2);
    \coordinate (antiquark) at (4.5,-1.2);

    \fill[red] (3.0,-1.1) rectangle (3.2,1.1);

    \draw[gluon line] (incoming) -- (splitting);
    \draw[fermion line] (splitting) -- (quark);
    \draw[fermion line] (antiquark) -- (splitting);

    \node[above=5pt] at (incoming) {$p,\mathbf{x}_0$};
    \node[above right=2pt] at (quark) {$p_1,\mathbf{x}_1$};
    \node[below right=2pt] at (antiquark) {$p_2,\mathbf{x}_2$};
\end{tikzpicture}